\documentclass[twocolumn]{article}

\usepackage{etex}
\reserveinserts{28}
\usepackage[natbib=true, autocite=superscript, style=nature, sorting=none, isbn=false, backend=bibtex8]{biblatex}

\usepackage{mathtools}
\usepackage{textcomp}
\usepackage{geometry}
\usepackage{authblk} 
\usepackage{amsmath}
\usepackage{amssymb}
\usepackage{bm}
\usepackage{nicefrac}
\usepackage{amsfonts}
\usepackage{mathrsfs}
\usepackage{hyperref}

\usepackage{graphics}
\usepackage[outdir=./]{epstopdf}
\usepackage{floatrow}
\usepackage[labelfont=bf]{caption}

\usepackage[switch]{lineno}
    
\usepackage{color}
\usepackage[markup=nocolor]{changes}

\title{Light controls motility and phase separation of photosynthetic microbes}

\author[1]{Alexandros A. Fragkopoulos}
\author[1]{J\'{e}r\'{e}my Vachier}
\author[1]{Johannes Frey}
\author[1]{Flora-Maud Le Menn}
\author[1]{Michael Wilczek}
\author[2,1]{Marco G. Mazza}
\author[1]{Oliver B\"{a}umchen}

\affil[1]{Max Planck Institute for Dynamics and Self-Organization (MPIDS), Am Fa{\ss}berg 17, D-37077 G\"{o}ttingen, Germany}
\affil[2]{Interdisciplinary Centre for Mathematical Modelling and Department of Mathematical Sciences, Loughborough University, Loughborough, Leicestershire LE11 3TU, United Kingdom} 
\affil[*]{To whom correspondence should be addressed. E-mail:\ oliver.baeumchen@ds.mpg.de}

\date{\today}

\definecolor{darkgreen}{rgb}{0, 0.5, 0}

\begin{document}
\maketitle

\begin{refsegment}

\noindent
\textbf{Large ensembles of interacting, out-of-equilibrium agents are a paradigm of active matter. 
Their constituents' intrinsic activity may entail the spontaneous separation into localized phases of high and low densities \autocite{buttinoni2013dynamical,zottl2014hydrodynamics,elgeti2015physics}.
Motile microbes, equipped with ATP-fueled engines, are prime examples of such phase-separating active matter \autocite{liu2013phase,liu2019self}, which is fundamental in myriad biological processes \autocite{simon2002microbial,dorkenJRSIn2012,costertonARMB1995,alldredgeDeepSea1989,kiorboeMarBio1990}.
The fact that spontaneous spatial aggregation is not widely recognized as a general feature of microbial communities challenges the generalization of phase separation beyond artificial active systems.
Here, we report on the phase separation of populations of \textit{Chlamydomonas reinhardtii} that can be controlled by light in a fully reversible manner.
We trace this phenomenon back to the light- and density-dependent motility, thus bridging the gap from light perception on the single-cell level to collective spatial self-organization into regions of high and low density.
Its spectral sensitivity suggests that microbial motility and phase separation are regulated by the activity of the photosynthetic machinery.
Characteristic fingerprints of the stability and dynamics of this active system paint a picture that cannot be reconciled with the current physical understanding of phase separation in artificial active matter, whereby collective behavior can emerge from inherent motility modulation in response to changing stimuli.
Our results therefore point towards the existence of a broader class of self-organization phenomena in living systems.}

%
\indent 
A common principle shared by active systems is the fact that the local behavior of the individual agent determines if and how the ensemble self-organizes into large-scale structures\autocite{elgeti2015physics}.
This local behavior is governed by the agent's intrinsic motility, as well as the mutual interactions between different agents.
Collective behavior of motile agents has been explored for a multitude of active systems, including populations of microscopic droplets\autocite{maass2016swimming}, colloids\autocite{golestanian2012collective,bauerle2018self,cao2019orientational} and  bacteria\autocite{sokolov2007concentration}.
Coherent self-organization may manifest itself with the local alignment of velocity\autocite{toner1995long,vicsek1995novel,cisneros2011dynamics} and also via the formation of clusters\autocite{buttinoni2013dynamical,zottl2014hydrodynamics,elgeti2015physics}.
The latter phenomenon has been described by a  positive feedback through a generic coupling of local density and velocity of physically interacting particles to explain the separation into high-density and low-density phases of colloids\autocite{cates2015motility,
FilyPRL2012,BialkePRL2012,redner2013structure,BialkeEPL2013,StenhammerPRL2013,WysockiEPL2014,barreJSP2015,pohlEPJE2015,digregorioPRL2018,solonPRE2018,barARCMP2019}
Models within this state-of-the-art theoretical framework are based on a density-dependent motility and successfully capture aggregation of artificial active particles in simulations and experiments\autocite{buttinoni2013dynamical,redner2013structure,levis2014clustering}. 
Motile microorganisms are intrinsically far from equilibrium and may in principle fulfill the same basic prerequisites as artificial agents that interact through steric and hydrodynamic interactions. 
However, they may also regulate their motility according to the density of neighbours (quorum sensing), respond to external chemical (chemotaxis) and light (phototaxis) gradients and be subject to birth and death. 
In light of this rich behavior, the generalization of theoretical frameworks towards microbial populations requires full external control of the motility characteristics in the absence of any effects other than physical interactions of the living species.
Generally, the investigation of the principles guiding spontaneous aggregation in microbial communities has remained scarce, which is rather surprising because of its enormous relevance for the marine food web\autocite{valiela1995marine,kiorboeMarBio1990,simon2002microbial,burd2009particle}, and biofilms\autocite{costertonARMB1995}.

\indent 
In the present work, we demonstrate that motility and spatial phase separation in active suspensions of \textit{Chlamydomonas reinhardtii} cells in a confined environment can be directly controlled by light. 
In the absence of photo-, chemo- and gravitaxis, we show that these light-perceiving microbes exhibit a light- and density-dependent motility, which allows to traverse into a regime of spontaneous phase separation of the active suspension. 
In this regime, the velocity and diffusivity of the motile cells exhibit power-law behaviors that are at variance with predictions from motility-induced phase separation (MIPS)\autocite{cates2015motility}.
Interestingly, the appearance as well as the dynamics of phase separation are governed by the light intensity and wavelength as control parameters, providing a direct link between the activity of the photosynthetic machinery, microbial motility and large-scale self-organization.

\begin{figure*}
\includegraphics{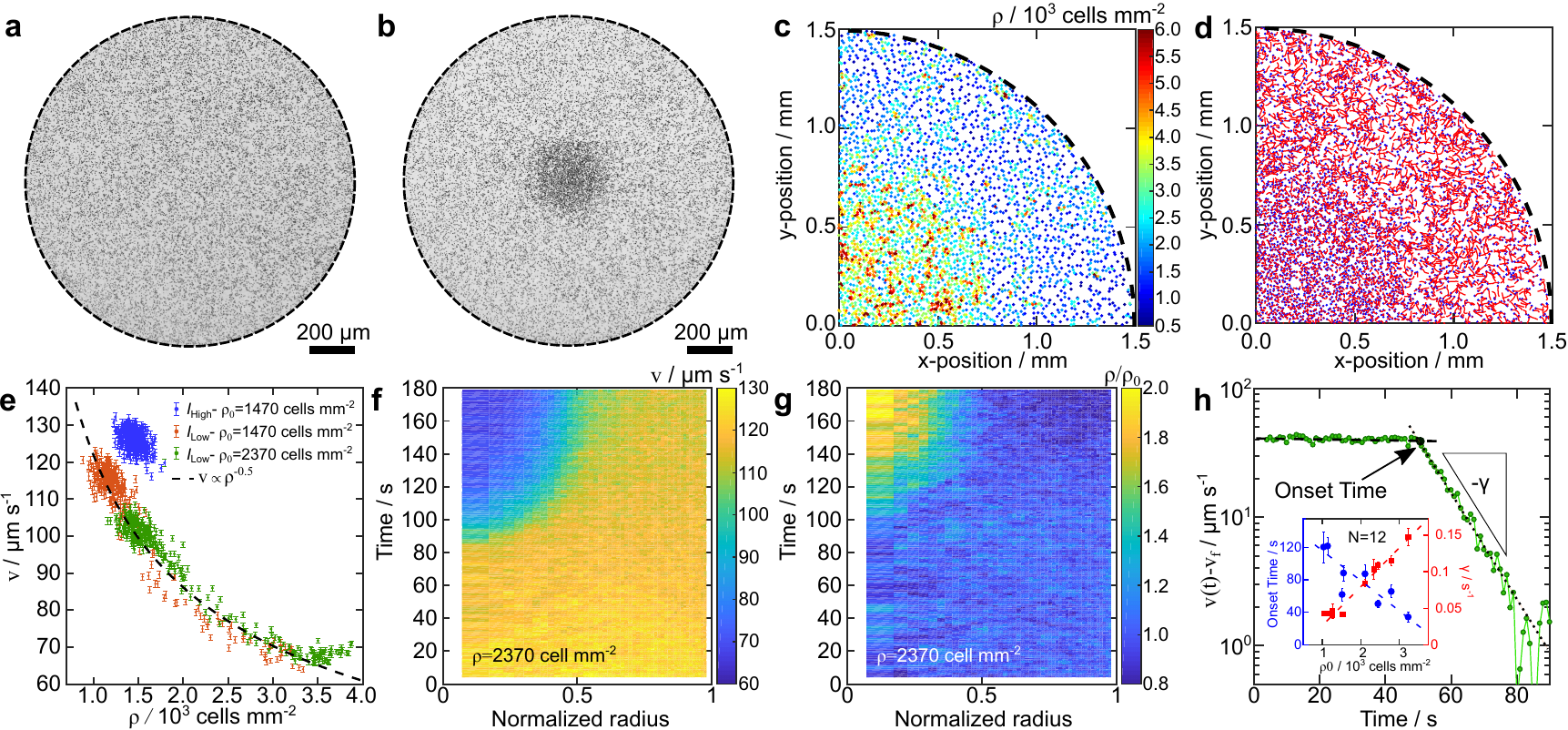}
\caption{\textbf{Light-switchable phase separation of confined \textit{C.~reinhardtii} suspensions.} 
Freely swimming \textit{C.~reinhardtii} cells are confined in a cylindrical compartment (radius $R=1.5$\,mm, height $h=21$\,\textmu m) and illuminated with red light ($\lambda = 671\pm5$\,nm). 
\textbf{a},\textbf{b} A cell suspension of global cell density $\rho_0 = 1300$\,cells mm$^{-2}$ is homogeneous (\textbf{a}) for a light intensity $I=20$\,\textmu mol m$^{-2}$ s$^{-1}$ (referred to as high), and separated into a dense phase in the center and a dilute phase otherwise (\textbf{b}) for $I=0.37$\,\textmu mol m$^{-2}$ s$^{-1}$ (referred to as low). 
\textbf{c},\textbf{d} Cell tracking analysis for an experiment at $I=0.37$\,\textmu mol m$^{-2}$ s$^{-1}$ and $\rho_0 = 2370$\,cells mm$^{-2}$. 
\textbf{c} Snapshot with cell positions color-coded according their local cell density $\rho$ measured by Voronoi tessellation. 
\textbf{d} Snapshot with cell positions (blue dots) and their trajectories (red lines) over the previous $0.5$\,s.
\textbf{e} Velocity-density coupling $v(\rho)$ for different light intensities $I$ and global cell densities $\rho_0$.
The dashed line corresponds to $v\propto\rho^{-0.5}$.
\textbf{f},\textbf{g} Emergence of phase separation as evidenced by the spatio-temporal evolution of $v$ and $\rho$, after switching to low light intensity at time $t$ = 0\,s. 
\textbf{h} Semi-log representation of the cell velocity $v(t)$ at the center of the compartment, where $v_\mathrm{f}$ denotes the steady state velocity within the dense phase. 
Lines indicate best linear fits to the experimental data before (dashed) and after (dotted) the onset of phase separation; their intersect determines the onset time $t_\mathrm{on}$. 
The inset displays onset time $t_\mathrm{on}$ and velocity decay rate $\gamma$ as a function of $\rho_0$; dashed lines represent best linear fits to the data. 
$N$ indicates the number of independent experiments.
}
\label{fig:fig1}
\end{figure*}

\paragraph*{Aggregation of motile cells at low light intensity}\mbox{}\\
\noindent In our experimental setup, a suspension of motile \textit{C.~reinhardtii} cells is confined in a circular compartment of radius $R=1.5\pm0.05$\,mm and height $h=21\pm1$\,\textmu m, as displayed in Fig.~\ref{fig:fig1}a,b. 
This quasi-2D confinement allows for identifying and tracking a large number of cells in their planktonic, i.e.\ free swimming, state over extended time periods. 
Experiments are performed using bright field microscopy and red light  (wavelength $\lambda = 671\pm6$ nm) in order to safely inhibit phototactic effects \autocite{berthold2008channelrhodopsin} and surface attachment of the cells \autocite{kreis2018adhesion}. 
When a dense suspension of \textit{C.~reinhardtii} cells is exposed to a light intensity $I=20$\,\textmu mol m$^{-2}$ s$^{-1}$, in the following referred to as ``high'' light intensity, we find that the cells are homogeneously distributed in the compartment (see Fig.~\ref{fig:fig1}a). 
Remarkably, when the light intensity is decreased to $I=0.37$\,\textmu mol m$^{-2}$ s$^{-1}$, referred to as ``low'' light intensity, the same community of motile cells exhibits a spatially inhomogeneous distribution featuring a substantially enhanced cell density in the center of the compartment, see Fig.~\ref{fig:fig1}b. This effect is completely reversible when the light condition is reverted (see Supplementary Fig.~1).

\indent 
Besides spatially localized density variations (Fig.~\ref{fig:fig1}c), we find that also the cell motility exhibits spatial variations, with cells at high local cell densities moving slower than their counterparts at lower cells densities (Fig.~\ref{fig:fig1}d), which points at a collective effect mediated by steric and hydrodynamic interactions.
A generic coupling of swimming motility and local density becomes evident by linking the local cell density, $\rho$, and the local root-mean-square (rms) speed of the cells, $v$.
At high light intensity, the $v$ vs.~$\rho$ scatter plot is highly localized exemplifying the fact that the system can be considered homogeneous (Fig.~\ref{fig:fig1}e). 
In contrast, we observe a bimodal distribution of velocities and densities for low light intensity, representing the dense phase in the center and the dilute phase elsewhere in the compartment (see Supplementary Fig.~2).
Notably, the transition between both phases follows a power-law with $v\propto \rho^{-0.5\pm0.1}$, independent on the global cell density of the suspension.
The direct comparison of the different steady states for the same cell density demonstrates that the rms velocity $v$ is substantially lowered at lower light intensities, which suggests that the emergence of the spatial pattern is induced by a modulation of the cell motility at different light intensities.

\begin{figure*}
\includegraphics{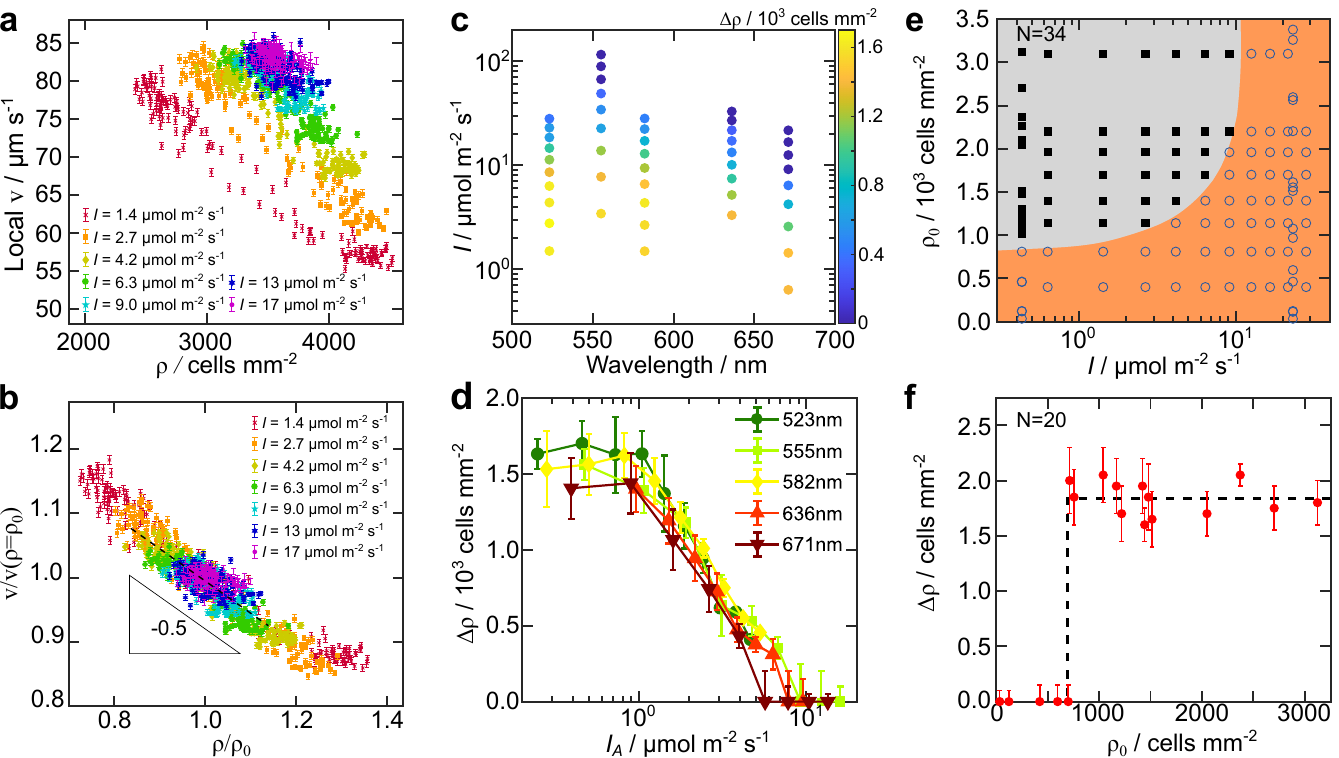}
\caption{\textbf{Light intensity and wavelength control the emergence of phase separation.} 
\textbf{a} Local rms cell velocity $v$ versus local cell density $\rho$ for different red light intensities ($\lambda=671$\,nm). 
\textbf{b} Scaled $\rho$ and $v$ of the experimental data shown in \textbf{a} using the global cell density, $\rho_0$, and the velocity at the global dell density, $v(\rho=\rho_0)$, respectively. 
\textbf{c} Color-coded representation of the order parameter, $\Delta\rho$, as a function of light intensity and wavelength. 
\textbf{d} Order parameter $\Delta\rho$ as a function of the chlorophyll-absorbed light intensity, $I_\mathrm{A}$, for different light wavelengths. 
All results shown in \textbf{a} to \textbf{d} were obtained from the same population of \textit{C.~reinhardtii} cell with $\rho_0=3400$\,cells mm$^{-2}$, whereby the light conditions were changed in random order. 
\textbf{e} Phase diagram $\rho_0$ versus $I$ characterizing the critical light intensity and critical cell density of phase separation (blue circles:\ $\Delta\rho=0$, black squares:\ $\Delta\rho\neq0$). 
\textbf{f} Above the critical cell density, $\Delta\rho$ is independent of the global cell density $\rho_0$ (shown for $I=0.37$\,\textmu mol m$^{-2}$ s$^{-1}$).
}
\label{fig:fig2}
\end{figure*}


\paragraph*{Phase separation kinetics}\mbox{}\\
\noindent To elucidate the phase separation phenomenon, we now turn to study its  dynamical properties. 
The local density $\rho$ and the swimming velocity $v$ are monitored as a function of both time and distance from the center of the compartment, see Fig.~\ref{fig:fig1}f,g, after switching the light intensity from high to low at time $t=0$. 
The rms swimming velocity $v$ is initially independent of the position within the compartment, as expected for the homogeneous state. 
From $t=0$ onward, we find that $v$ linearly decreases in the entire compartment at a comparably small constant rate of $0.08\pm0.03$\,\textmu m/s$^2$ (dashed line in Fig.~\ref{fig:fig1}h).
At a characteristic onset time $t_\mathrm{on}$ in the range of tens of seconds, the swimming velocity in the center of the compartment strongly decreases, see Fig.~\ref{fig:fig1}f.
This high-density region continuously grows in size until after a few minutes it reaches a steady state, characterized by a well-defined final size (see Supplementary Fig.~3).
In accordance with the previously described coupling $v(\rho)$ for the steady states, also the local cell density increases in the center of the compartment, with a time lag of about 4 to 7\,s to $t_\mathrm{on}$, see Fig.~\ref{fig:fig1}g.
This sequence of events suggests that the change in light intensity has an immediate effect on the motility of the cells, which subsequently, via the generic $v(\rho)$ coupling, manifests in changes of the cell density.

\indent 
The kinetics of the phase separation process are characterized by the local velocity in the center of the compartment decreasing exponentially with time, i.e.\ $v-v_f\propto e^{-\gamma t}$, where $\gamma$ is the velocity decay rate and $v_f$ the steady state velocity within the dense phase, see Fig.~\ref{fig:fig1}h.
Both the onset time $t_\mathrm{on}$ and the decay rate $\gamma$ decrease linearly with the global cell density, see inset of Fig.~\ref{fig:fig1}h, i.e.\ phase separation sets in earlier and dense and dilute phases form faster at higher cell densities.

\paragraph*{Light intensity and color discrimination}\mbox{}\\
\noindent The emergence of the phase separation in \textit{C.~reinhardtii} suspensions is directly linked to the motility characteristics of the individual planktonic cells, which are considerably affected by the light intensity. 
The motility of each individual microbe is driven by its continuous and coordinated flagella beating, which is fueled by the conversion of ATP into mechanical stresses by the molecular motors at play \autocite{wan2016coordinated}.
\textit{C.~reinhardtii} is a mixotrophic microorganism, and as such it is able to produce ATP either through photosynthesis, and through the respiratory system via the consumption of sugars. 
By reducing the light intensity, the photosynthetic activity of the cells is lowered, eventually forcing them to switch from using the photosynthetic mechanism and to respiration \autocite{xue1996interactions}. 

\indent 
In the following, we establish the link between light-regulated energy conversion, motility characteristics, and phase separation in photoactive suspensions.
So far, we reported on active suspensions that were illuminated with a single wavelength $\lambda=671$\,nm and two light intensities, which we referred to as ``low'' and ``high'', see Fig.~\ref{fig:fig1}.
For the exact same microbial suspension, we now systematically tune the light intensity and wavelength, and evaluate the distributions of local velocity $v$ and cell density $\rho$, see Fig.~\ref{fig:fig2}.
We find that, below a critical light intensity, the suspension gradually transitions from a homogeneous state to a phase-separated state. 
Upon decreasing the light intensity, the initial unimodal distribution of data points splits into a bimodal distribution, where the two clusters, corresponding to the dilute and the dense phase, progressively move apart, see Fig.~\ref{fig:fig2}a. 
In addition, the overall velocity of the cells monotonically decreases with decreasing light intensity, exemplifying that the light intensity can be used as a control parameter for the cell motility. 
We find that a cumulative plot of the normalized velocity, $\tilde{v} = v/v(\rho=\rho_0)$, versus the normalized density, $\tilde{\rho}=\rho/\rho_0$, collapses all light intensities onto one master curve with a slope of $-0.5\pm0.1$. 
This implies $\left.d v(\rho)/d\rho\right|_{\rho=\rho_0}=-0.5 v(\rho_0)/ \rho_0$, which is in accordance with the power-law dependence $v\propto \rho^{-0.5}$ shown in Fig.~\ref{fig:fig1}e, and demonstrates that the same power-law persists for all different light intensities (see Fig.~\ref{fig:fig2}b) and global densities (see Supplementary Fig.~4).

\indent 
In analogy to gas/liquid phase separation, we further introduce the difference between the average cell densities in the dense and dilute phase $\Delta\rho$ as an order parameter \autocite{beysens1987phase} (see Supplementary Fig.~2a). 
Fig.~\ref{fig:fig2}c displays $\Delta\rho$ for different light intensities and wavelengths. 
Clearly, $\Delta\rho$ does not only depend on the light intensity, as evident from Fig.~\ref{fig:fig2}a,b, but also the wavelength of the illumination controls $\Delta\rho$.
As shown in Fig.~\ref{fig:fig2}c, the critical light intensity for phase separation is about one order of magnitude smaller for red ($\lambda=671$\,nm) compared to green light ($\lambda=555$\,nm), suggesting that chlorophyll a and b, which are responsible for the regulation of the photosynthetic machinery and the synthesis of ATP, might be involved.
Indeed, we find that $\Delta\rho$ versus the light intensity $I_\mathrm{A}$ absorbed by chlorophyll for each wavelength\autocite{kan1976light} collapse onto a master curve, see Fig.~\ref{fig:fig2}d, suggesting that the regulation of the photosynthetic machinery governs phase separation of \textit{C.~reinhardtii} suspensions.
For decreasing $I_\mathrm{A}$, the order parameter $\Delta\rho$ increases and saturates at about 1500\,cells mm$^{-2}$ for low light intensities.

\begin{figure*}
\includegraphics{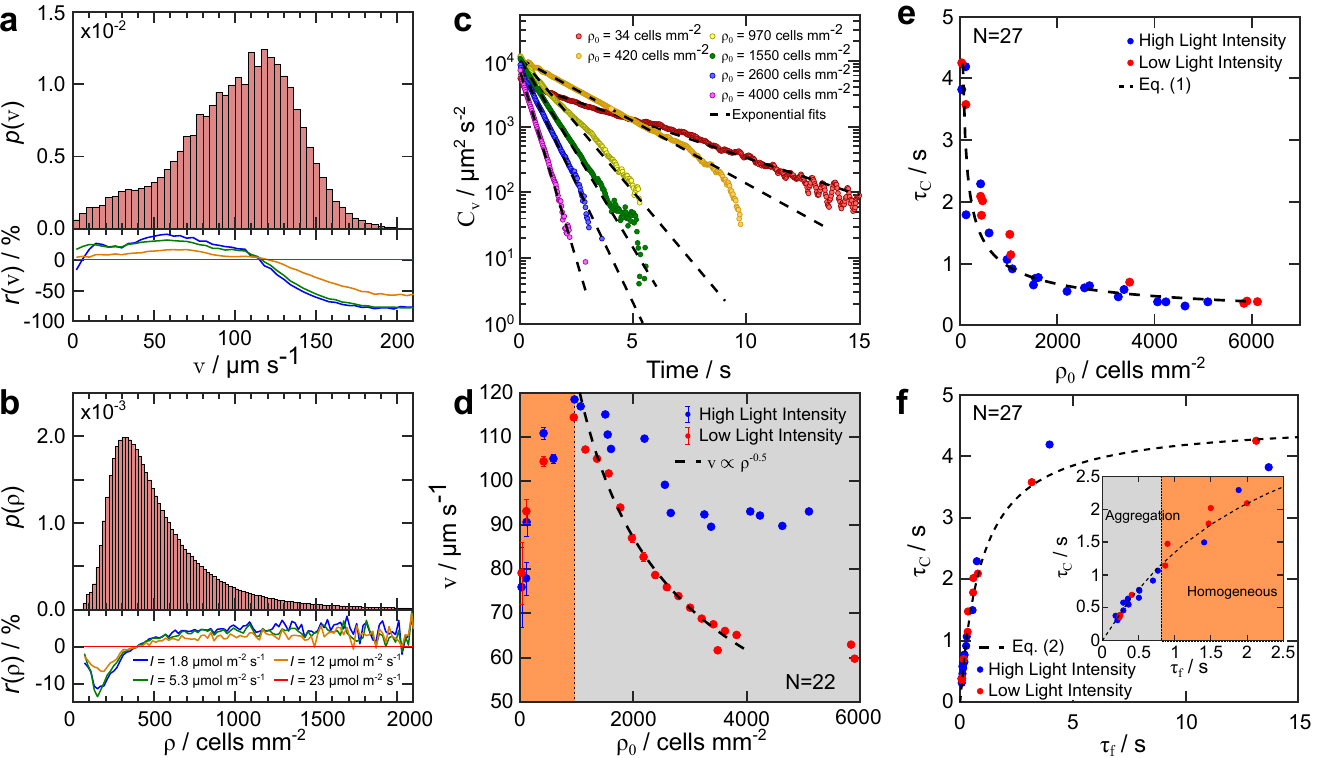}
{\caption{\textbf{Motility characterization of dilute and dense \textit{C.~reinhardtii} suspensions.} 
\textbf{a},\textbf{b} Motility of a dilute suspension with $\rho_0=400$\,cells mm$^{-2}$ for different light intensities. 
\textbf{a} displays the probability distribution of velocities, $p(v)$, for the light intensity $I=20$\,\textmu mol m$^{-2}$ s$^{-1}$, while \textbf{b} shows the corresponding probability distribution of local densities $p(\rho)$.
The residuals $r(v)$ and $r(\rho)$ indicate how the distributions $p(v)$ and $p(\rho)$ change by reducing the light intensity. 
\textbf{c} Semi-log representation of the velocity autocorrelation function, $C_v$, for $I=20$\,\textmu mol m$^{-2}$ s$^{-1}$ and different $\rho_0$. 
The dashed lines correspond to best linear fits to the data. 
\textbf{d} Root-mean-square velocity as a function of local cell density for $I_{\mathrm{low}}=0.37$\,\textmu mol m$^{-2}$ s$^{-1}$ and $I_{\mathrm{high}}=20$\,\textmu mol m$^{-2}$ s$^{-1}$. 
The dotted line indicates the critical cell density $\rho_c$. 
\textbf{e} Velocity correlation time, $\tau_C$, as a function of the local cell density at $I_{\mathrm{low}}$ and $I_{\mathrm{high}}$. 
The dashed line corresponds to $\tau_C\propto\rho^{-0.5}$. 
\textbf{f} $\tau_C$ as a function of the collision time, $\tau_{f}$. 
The dashed line is a best fit to Eq.~(\ref{eq:eq1}). 
Inset: Close-up of the same data at the transition to phase separation. 
Phase separation occurs only in the linear regime, i.e.\ $\tau_C\propto\tau_f$, where cell-cell interactions dominate. 
$N$ is the number of independent experiments.}
\label{fig:fig3}}
\end{figure*}

\indent 
Aside from the light intensity and wavelength, also the global cell density plays a key role on the appearance of phase separation. 
We explore the phase space of the phenomenon by gradually varying the light intensity for active suspensions of different cell densities and summarize the results in a phase diagram, see Fig.~\ref{fig:fig2}e. 
Interestingly, we find that there is not only a critical light intensity for the phase separation to appear, but also a critical cell density, $\rho_c$, at about 800\,cells\,mm$^{-2}$ for a light intensity $I=0.37$\,\textmu mol m$^{-2}$ s$^{-1}$. 
As shown in Fig.~\ref{fig:fig2}f, we find that there is a discontinuity in $\Delta\rho$ at the critical cell density $\rho_c$, which is strikingly different to the continuous transition observed for experiments at constant $\rho_0$, see Fig.~\ref{fig:fig2}d. 
Taking these findings together, we revealed a complex interplay of light conditions and cell density regulating the emergence as well as the appearance of phase separation of confined \textit{C.~reinhardtii} suspensions. 
Unraveling the mechanism at the cellular level underlying the emergence of phase separation of motile microbes ultimately requires establishing a quantitative link between light conditions, cell density and microbial motility, which we address in the following.


\paragraph*{Unraveling the motility--density dependence}\mbox{}\\
\noindent We first consider cell densities $\rho_0<\rho_c$, where the confined suspension is spatially and temporarily homogeneous for all light intensities. 
In this regime, the probability distributions of velocity $p(v)$ and local density $p(\rho)$, respectively, exhibit a single maximum, as shown in Fig.~\ref{fig:fig3}a,b.
Upon decreasing the light intensity for the same cell suspension, we identify characteristic deviations of $p(v)$ and $p(\rho)$.
These changes are quantified using the residual $r(v) = \left(p(v)-p_h(v)\right)/p_h(v)$, where $p_h(v)$ is the velocity distribution at high light intensity. 
As shown in Fig.~\ref{fig:fig3}a, the probability of finding smaller cell velocities gradually increases with decreasing light intensity.
As a result of the velocity-density coupling, see Fig.~\ref{fig:fig1}e, the variation of $p(v)$ causes a corresponding variation of $p(\rho)$, which manifests as a higher probability of finding enhanced local cell densities at lower light intensities, see Fig.~\ref{fig:fig3}b.
Interestingly, neither the values nor the positions of the maxima of $p(v)$ and $p(\rho)$ are affected by changing the light intensity.

\indent Having established the link between light intensity and cell motility, we now explore the coupling between cell motility and density on the basis of quantitative statistical measures of the experimental data and put these measures into perspective of the theoretical frameworks of active Brownian (AB) and run-and-tumble (RT) motion.
Previous studies of the motility of \textit{C.~reinhardtii} were exclusively limited to single cells, showing that the cell's motion is predominantly ballistic on short time scales and exhibits sudden reorientations at a typical time $\tau_R$ \autocite{polin2009chlamydomonas,ostapenko2018curvature}, a behavior that is described as RT motion.
We quantify the cell motility in dense \textit{C.~reinhardtii} suspensions using the velocity autocorrelation function \autocite{alder1967velocity}, $C_v(t) = \langle \textbf{v}_i(t=0)\cdot\textbf{v}_i(t)\rangle$, where $\textbf{v}$ is the velocity of an individual cell $i$.
The average is taken over all tracked cells. 
As shown in Fig.~\ref{fig:fig3}c, $C_v$ follows an exponential decay, which is also expected for ensembles of AB particles \autocite{zeitz2017active}.
We thus express the velocity autocorrelation function as $C_v=v^2e^{-\nicefrac{t}{\tau_C}}$, where $v=\sqrt{\langle \textbf{v}^2 \rangle}$ is the root-mean-square velocity, and $\tau_C$ denotes the velocity correlation time.
For both RT and AB motion\autocite{taktikos2013motility}, $\tau_C$ characterizes particle reorientations and is linked to the rotational diffusion constant $D_\mathrm{R}=(2\tau_C)^{-1}$.

\indent 
In the following, we quantify $v$ as a population average for the experimental data at different global densities $\rho_0$ and two different light intensities.
As shown in Fig.~\ref{fig:fig3}d, $v$ exhibits a non-monotonic behavior for both light intensities:
The velocity initially increases with increasing $\rho$, reaches a maximum of about $120$\,\textmu m/s at a density of about 800\,cells/mm$^2$, and decreases for larger $\rho$ approaching different minimum levels depending on the light intensity, i.e.\ $91\pm2$\,\textmu m/s for high light intensity and $63\pm3$\,\textmu m/s for low light intensity.
In particular, the decrease of $v(\rho)$ in the case of the low light intensity follows the power-law $v\propto \rho^{-0.5}$, in agreement with the velocity-density distribution inferred from a single phase-separated population, see Fig.~\ref{fig:fig1}e.
The non-monotonic behavior $v(\rho)$ suggests a competition of two counteracting mechanisms: 
For $\rho<\rho_c$, where the critical density $\rho_c$ represents the position of the maximum, the swimming cells predominantly interact via hydrodynamic coupling of their local flow fields causing an effective enhancement of the average motility\autocite{volpe2014simulation}.
For sufficiently large densities $\rho>\rho_c$, however, the increased frequency of collisions causes steric interactions to dominate over hydrodynamic interactions\autocite{matas2014hydrodynamic}.
From simulations in quasi-2D\autocite{theers2018clustering}, the shape of $v(\rho)$ is expected to be linear for purely steric interaction, but becomes similar to the one observed in our experiments after including hydrodynamics.
The value of $\rho_c$ matches the critical cell concentration above which  phase separation is observed, see Fig.~\ref{fig:fig2}f.
Thus, $v(\rho)$ being a monotonically decreasing function appears to be an essential precondition for the emergence of phase separation into localized dense and dilute phases of motile cells.

\indent 
The mechanism underlying the generic velocity-density coupling are frequent collisions of the swimming cells, which we characterize by evaluating the velocity correlation time $\tau_C$ for different global densities $\rho_0$.
For both light intensities, $\tau_C$ is decreasing for increasing density, see Fig.~\ref{fig:fig3}e, i.e.\ the velocity of the cells decorrelates faster for increasing $\rho$, which results from more frequent reorientations of the motile cells due to more frequent cell-cell interactions.
In the framework of ensembles of AB particles in 2D, $\tau_C\propto\tau_f$ is predicted, where $\tau_f$ is the time between collisions.
We now write the collision time as $\tau_{f} = l_{f}/v$, where $l_{f}=(\sqrt{8} d\rho)^{-1}$ is the mean-free-path in 2D and $d$ the cell diameter \autocite{tong2012kinetic}.
From $\tau_C \propto l_{f}/v$ and $l_{f}\propto \rho^{-1}$, together with the generic coupling $v\propto \rho^{-0.5}$ extracted from phase separation experiments (see Fig.~\ref{fig:fig1}e and Fig.~\ref{fig:fig2}b), we expect
\begin{equation}\label{eq:eq0}
\tau_C\propto\rho^{-0.5}, 
\end{equation}
which is validated by our motility analysis (Fig.~\ref{fig:fig3}e, dashed line).
For individual \textit{C.~reinhardtii} cells, however, the motility is governed by their RT motion\autocite{polin2009chlamydomonas}, featuring a characteristic tumble time $\tau_R$.
Thus, we expect $\tau_C = \tau_R$ in the density limit where the velocity decorrelates due to tumble events rather than cell-cell collisions. 
By adding the rotational diffusion constant for the collision-dominated and the tumble-dominated regime yields $1/\tau_C = 1/(c\tau_f)+1/\tau_R$, where $c$ is the number of collisions needed for the velocity to decorrelate.
Thus, we find 
\begin{equation}\label{eq:eq1}
    \tau_C=\frac{c \tau_f \tau_R}{c \tau_f+\tau_R},
\end{equation}
which quantitatively captures the experimental data using $v$, $\rho$ and $d$ from $N$ independent experiments, see Fig.~\ref{fig:fig3}f.
The values of $\tau_C$ increase linearly with $\tau_f$ in the collision-dominated limit, i.e.\ for high cell densities and small $\tau_f$, and saturate for large $\tau_f$ to a constant value representing the tumble time of a single cell\autocite{polin2009chlamydomonas}. 
A best fit of Eq.~(\ref{eq:eq1}) to the data yields $c=4.8\pm0.2$ and $\tau_R=4.2\pm0.3$\,s. 
By writing the rotational diffusion constant for the collision-dominated regime as $D_{R,c}=\Delta\theta^2/2\tau_f$, we find that $\Delta\theta$, the change of the orientation of a cell after a collision, remains on average constant and has a value of $\Delta\theta=26^{\circ}$. 
Spontaneous phase separation occurs for $\tau_{f}<0.8$\,s, which coincides with the regime where $\tau_C$ is linear with $\tau_f$ and hence when the motility of the cells is dominated by cell-cell collisions, as shown in the inset of Fig.~\ref{fig:fig3}f.


\indent 
Phase separation in this active living system is exclusively controlled by physical interactions, in the absence of any external stimuli (see Supplementary Figs.~5).
Besides cell-cell interactions as discussed above, these also include cell-wall encounters due to the constraints of the habitat.
Implementing physical boundaries, and thereby steric cell-wall interactions, for a many-particle system of AB agents reveals that they control the characteristic appearance of the phase-separated state (see Supplementary Fig.~6), including the localization of the dense phase in the center of the compartment (see Supplementary Fig.~7).
Active cell-cell and passive cell-wall interactions, though both predominantly steric at high cell density, are fundamentally different on the single-cell level, which appears to be a key ingredient of the symmetry-breaking process.


\paragraph*{Discussion}\mbox{}\\
\noindent 
Taking together all of the aforementioned measurements reveal a complex interplay of the density of motile cells, their motility characteristics and the light conditions, which allows for isolating the key requirements of phase separation of an ensemble of motile photoactive microbes:
First, when the active suspension is sufficiently dense cell-cell interactions dominate the reorientation due to natural tumbling motion.
Second, the local velocity develops an inverse power-law relationship with the local density.
This feature is an essential ingredient of a positive feedback mechanism that allows for phase separation to develop, since the cell flux, $\rho \textbf{v}$, between the dense and dilute phase has to be constant.
This aspect of the phenomenon shares conceptual similarities with motility-induced phase separation (MIPS), where the inhibition of a particle's ability to reorient and escape nearby particles result in particle clustering \autocite{suma2014motility,cates2015motility,patch2017kinetics}. 
However, phase separation of dense suspensions of motile \textit{C.~reinhardtii} cells is at variance with predictions from MIPS, which requires the exponent of $v\propto \rho$ to be smaller than -1.
\textit{C.~reinhardtii} cells are not trapped only due to difficulty reorienting, but also due to the decrease in the velocity at low light intensity.
The light response is a unique effect of the photosynthetic activity, which we identified as key parameter governing motility and phase separation of \textit{C.~reinhardtii} suspensions.
This phase separation in confinement is solely based on physical interactions of light-perceiving cells and occurs in the complete absence of phototaxis, chemotaxis, gravitaxis and quorum sensing.

\printbibliography[notkeyword=appendix]
\end{refsegment}


\paragraph{Acknowledgements}
The authors thank the G\"ottingen Algae Culture Collection (SAG) for providing the \textit{Chlamydomonas reinhardtii} strain SAG~11-32b. 
We thank M.\ Lorenz for fruitful discussions and technical assistance.

\paragraph{Author Contributions}
O.B.\ and A.A.F.\ designed research. 
A.A.F.\ led the experiments and A.A.F., J.F., and F.-M.L.\ performed them.
M.W.\, M.G.M.\ and A.A.F.\ led the theoretical analysis. 
M.G.M\ and J.V.\ led the simulations and J.V.\ performed them. 
A.A.F., J.V., J.F., M.W., M.G.M\ and O.B.\ contributed to the interpretation of the data. 
A.A.F. wrote the first draft of the manuscript. 
All authors contributed to the discussions and the final version of the manuscript.


\renewcommand{\figurename}{Supplemental Material, Figure}
\setcounter{figure}{0}

\newcommand{\beginsupplement}{%
        \setcounter{table}{0}
        \renewcommand{\thetable}{\arabic{table}}%
        \setcounter{figure}{0}
        \renewcommand{\thefigure}{\arabic{figure}}%
     }
\beginsupplement

\newpage
\clearpage

\begin{refsegment}

\section*{Methods}

\paragraph*{Cell cultivation}
Wild-type \textit{C.~reinhardtii} cells, strain SAG~11-32b, were grown axenically in Tris-Acetate-Phosphate (TAP) medium on 12\,h$-$12\,h day-night cycles at temperatures of $24\,^\circ$C and $22\,^\circ$C, respectively, in a Memmert IPP 100Plus incubator. 
The daytime light intensity was $28\pm10$ \textmu mol m$^{-2}$ s$^{-1}$, while during the nighttime the light intensity was reduced to zero. 
Experiments were performed with vegetative cells taken from the cultures in logarithmic growth phase during the daytime on the third day after incubation.
The experimental culture was centrifuged for $10$\,min at an acceleration of $100$\,g, the excess fluid was removed, and the pellet of cells was resuspended in fresh TAP medium. 
Since cells may deflagellate due to the mechanical shearing during the centrifugation, the cell suspension was allowed to rest for $2$\,hours after  centrifugation, which is sufficient to regrow their flagella \autocite{harris2009chlamydomonas}.
In order to control the overall cell concentration in the suspension, a hemocytometer (Neubauer-improved with double net ruling) was used for manual cell counting. 
The final cell suspension density before experiments ranged between $1\times10^6$ to $8\times10^7$\,cells/mL.

\paragraph*{Experimental setup}
All experiments were performed with an Olympus IX83 inverted microscope and combining LED illumination (CoolLED) with bandpass filters. 
The condenser was always adjusted to ensure K\"{o}hler illumination, which provides homogeneous and reproducible light conditions. 
A 4x objective was used in conjunction with the magnification changer set at 1.6x resulting to a total of 6.4x magnification. 
This configuration provides a large field of view that includes the majority of the compartment (1.95\,mm $\times$ 1.95\,mm), while it is large enough to perform single-cell tracking.

The compartments for the cell suspensions were made out of polydimethylsiloxane (PDMS). 
Specifically, the commercially available Sylgard 184 PDMS was used in all experiments, and the mixing ratio used to produce the PDMS is 10:1 by weight base to curing agent for all experiments. 
In order to achieve the desired compartment height, the mixture of PDMS was spin coated on a glass slide at $1300$\,rpm for $5$\,minutes, to achieve the final thickness of $21\pm1$\,\textmu m. 
The glass slide was immediately placed on a hot plate at about $95\,^\circ$C to accelerate the cross-linking process of the PDMS chains. 
Afterwards, the glass slides were placed in an oven at $75^\circ$C for 2\,hours to complete the polymerization process, and a circular punch of $R=1.5$\,mm (Harris Uni-Core) was used to produce the desired circular compartment. 
For the experiments, $20$\,\textmu l of the suspension was pipetted in the compartment and it was sealed with a secondary glass slide. 

\paragraph*{Light conditions}
The light conditions were precisely controlled during all experiments.
All experiments regarding the phase separation dynamics and the motility characterization (Figs.~\ref{fig:fig1},\ref{fig:fig3}) were performed under red light to avoid both phototaxis \autocite{berthold2008channelrhodopsin} and adhesion to surfaces \autocite{kreis2018adhesion}. 
This was achieved using an interference bandpass filter with center wavelength of 671\,nm and a full width at which the intensity is half of the maximum (FWHM) of 10\,nm.
For the color discrimination experiments (Fig.~\ref{fig:fig2}), additional interference bandpass filters were used: 523\,nm (FWHM:\ 23\,nm), 555\,nm (FWHM:\ 30\,nm), 582\,nm (FWHM:\ 10\,nm), and 636\,nm (FWHM:\ 20\,nm). 
Wavelengths below $523$\,nm were not considered since the cells may exhibit both phototaxis \autocite{berthold2008channelrhodopsin} surface adhesion \autocite{kreis2018adhesion}.
All light intensities were measured using a Thorlabs PM100D powermeter (Thorlabs S130C photodiode power sensor) for monochromatic light.

\paragraph*{Data recording and cell tracking}
All data were recorded using a Photometrics Iris 9 camera at $33$\,fps.
The combination of the relatively high quantum efficiency ($>73$\%) together with the 16-bit depth allowed for recording data sufficiently fast at the lowest light intensities. 
The array size of 2960\,px$\times$2960\,px allowed for imaging a large field of view to perform the experiments.
A background image for each experiment was calculated as the maximum intensity projection of the image sequence. 
The maximum intensity was selected since cells appear darker in the image. 
Hence, only stationary cells and objects will appear dark in the background image. 
The background image was then subtracted from the image sequence to remove detection of any stationary objects. 
For very low light intensities, there are intensity fluctuations within the image sequence that can diminish the particle detection. 
We solve this issue by scaling the intensity of each image such that the average is constant for each image sequence. 
Finally, we applied a 2.7$\times$2.7\,\textmu m median filter.

The cells were detected using the circular Hough transform \autocite{pedersen2007circular}, which only needs a part of the boundary of a circle to detect it. 
This is important in our case since we lose part of the cell edges when the density becomes sufficiently large. 
The disadvantage of the circular Hough transform is that it requires a priori knowledge of the expected radius of the objects in the image. In our case we used a range of radii between $3.3$\,\textmu m and $7.2$\,\textmu m. 
Once the cells have been detected, their trajectories are determined by using a Matlab based code by Blair and Dufresne \autocite{blair2008matlab}. 
The only parameter needed for the calculation is the maximum distance a particle can travel between frames. 
We chose this parameter by setting the maximum possible velocity of the cells to be $200$\,\textmu m/s. 

The cell velocity is taken as the displacement of the tracked cells, $\Delta x$, over one frame, $\Delta t=30$\,ms, and, thus, defined as $v = \Delta x/\Delta t$. 
The local cell density $\rho$ is calculated using the Voronoi tessellation of the tracked cells in each frame. 
Voronoi tessellation partitions the space into regions, with each region associated to a single cell. 
In addition, any point in a given region is closer to the associated cell of that regions compared to any other cell. 
We define the area of the region associated to $i$th tracked cell as $A_i$, and the corresponding local density as $\rho_i=A_i^{-1}$. 
A distribution of all $\rho_i$ is shown in Fig.~\ref{fig:fig3}b, while for the data shown in Fig.~\ref{fig:fig1}e and Figs.~\ref{fig:fig2}a,b, an azimuthal average has been performed. 

\paragraph*{Chlorophyll absorbance}
The absorbance of a substance is defined as $A = -\log_{10}T$, where $T=I_T/I_0$ is the transmittance with $I_T$ and $I_0$ the transmitted and incident light intensity. 
We define the absorbed intensity as $I_A = I_0-I_T = I_0\left(1-10^{-A}\right)$. 
Finally, we use the values of $A$ from literature for chlorophyll a and b obtained from \textit{C. reinhardtii} cells \autocite{kan1976light}.
By comparing the detected particles with the original images, we estimated an error in our cell detection. 
For $\rho<1500$ cells/mm$^2$, we have an error of about $2$\% in the density, which can become as large as $6$\% for the highest densities.

\paragraph*{Minimal active Brownian particle model}
We investigate a many-particle system composed of active Brownian particles under confinement in quasi-2D. Each particle is described by two overdamped Langevin equations 
\begin{align}
&\frac{d\bm{r}}{dt}=v\bm{e}-\nabla\phi_{\mathrm{WCA}}+\bm{\xi}\\
&\frac{d\bm{e}}{dt}=\bm{\xi_e}\times\bm{e}
\end{align}
where $\bm{r}$ and $\bm{e}$ are the position and orientation, respectively, of the particle, and $v$ its speed; the Gaussian noises $\bm{\xi}$ and $\bm{\xi_e}$ model the random fluctuations of the particle's position and the orientation, respectively. The particle-particle interaction is given by the Weeks--Chandler--Anderson potential \autocite{weeks1971role}, $\phi_{\mathrm{WCA}}$. 
The velocity depends on the density and is approximated by a step function. The maximum velocity is fixed according to the experimental value as well as the critical density, for which we observe the transition between the maximum and minimum velocities. 
The minimum velocity varies and represents the light intensity. Moreover, the translational and rotational diffusion coefficients depend on the local density.
The diffusion coefficients represent the thermal fluctuations in the surrounding medium but also the interactions with the other active particles \autocite{grossmann-NJP-2012,romanczuk-epjs-2012,lopez-pre-2006}. 
See the SI for more details.

\printbibliography[keyword=appendix]
\end{refsegment}

\end{document}